\documentclass[aps, prb, superscriptaddress, showpacs, floatfix, twocolumn]{revtex4-2}
\usepackage{graphicx}
\usepackage{hyperref}
\usepackage{amsmath}
\allowdisplaybreaks
\begin{document}
\title{Model studies of topological phase transitions in materials \\ with two types of magnetic atoms}
\author{Zhuoran He}
\affiliation{Wuhan National High Magnetic Field Center $\&$ School of Physics, Huazhong University of Science and Technology, Wuhan 430074, China}
\author{Gang Xu}
\email[Electronic address: ]{gangxu@hust.edu.cn}
\affiliation{Wuhan National High Magnetic Field Center $\&$ School of Physics, Huazhong University of Science and Technology, Wuhan 430074, China}

\begin{abstract}
\vspace*{0\baselineskip}
We study the topological phase transitions induced by Coulomb engineering in three triangular-lattice Hubbard models $AB_2$, $AC_3$ and $B_2C_3$, each of which consists of two types of magnetic atoms with opposite magnetic moments. The energy bands are calculated using the Schwinger boson method. We find that a topological phase transition can be triggered by the second-order (three-site) virtual processes between the two types of magnetic atoms, the strengths of which are controlled by the on-site Coulomb interaction $U$. This new class of topological phase transitions have been rarely studied and may be realized in a variety of real magnetic materials.
\end{abstract}

\pacs{}

\maketitle
\section{Introduction \label{sec:intro}}
Topological phase transitions \cite{PhysRevB.102.081110,PhysRevB.100.195432,PhysRevB.97.100302, PhysRevLett.119.187003} play a key role in condensed matter physics. Especially, magnetic topological systems \cite{PhysRevLett.125.117205,PhysRevX.9.041055,PhysRevLett.115.036805} often exhibit rich topological phases due the complicated interplay between electron-electron interactions, magnetic moments and spin-orbit coupling, which have been attracting intensive research interests for years \cite{PhysRevB.89.235416,PhysRevB.90.205111,PhysRevB.78.035123}. A general model for the description of magnetic topological insulators is the spin-orbit coupled Hubbard model \cite{PhysRevB.98.075117,PhysRevB.95.075124} with on-site Coulomb interaction $U$. Previous works on Coulomb engineering and correlation-driven effects in magnetic topological systems have studied various aspects of this topic including the Hartree-Fock mean-field theory \cite{PhysRevX.9.041055,Zhu_2019}, dynamical screening effects \cite{rosner2020coulombengineered}, and phase transitions due to the magnetic exchange coupling \cite{Lei_2007,doi:10.1143/JPSJ.58.1516,PhysRevB.40.2610} using the Schwinger boson method. These works mostly focus on systems with one type of magnetic atom, while the topological phase transitions in systems with two types magnetic atoms are comparatively less studied.

In this paper, we study systems with two types of magnetic atoms \cite{OHNISHI198355,MARCOS199016323,PhysRevLett.114.147204,PhysRevB.81.174402} with opposite magnetic moments. In such systems, the two types of magnetic atoms form two sets of Chern bands separately, which then interact via a type of second-order virtual process of order $\mathcal{O}(t_1t_2/U)$. These processes involve the hopping from one type of magnetic atom $i$ to atom $j$ via the other type of magnetic atom $k$ as an intermediate site. We call these $1/U$-controlled virtual processes the three-site terms, which can induce interesting topological phase transitions. We study their effect in a 2D hexagonal Hubbard model with 3 types of lattice sites $A, B, C$ forming triangular, honeycomb and Kagome sublattices, respectively. By putting spin-up and spin-down electrons on two of the three types of lattice sites, we consider $AB_2$, $AC_3$ and $B_2C_3$ models and realize $1/U$-controlled topological phase transitions as characterized by changes in the Chern numbers of the spin-up and spin-down bands. Our results demonstrate the interplay between band topology and correlation effects, and present Coulomb engineering as a powerful tool to manipulate the topological phases of matter, with potential applications in various solid-state physical systems.

The rest of the paper is organized as follows. In Sec.~\ref{sec:formalism}, we give the general formalism of our downfolding technique in the Schwinger boson representation and obtain the low-energy effective Hamiltonian containing the three-site terms. In Sec.~\ref{sec:results}, we apply our formalism to the $AB_2$, $AC_3$ and $B_2C_3$ lattice structures to demonstrate the $1/U$-controlled topological phase transitions. Section \ref{sec:conclusion} is a summary and conclusion, with discussions of potential materials to realize the topological phase transitions found in our model studies.

\section{Formalism \label{sec:formalism}}
Suppose an insulating magnetic material is described by the Hubbard model
\begin{align}
H=\sum_{ij\alpha\beta}t_{ij}^{\alpha\beta} c_{i\alpha}^\dagger c_{j\beta}+U\sum_i n_{i\uparrow} n_{i\downarrow},
\label{eq:Hubbard-model}
\end{align}
where $i,j$ are the site indices and $\alpha,\beta$ label the spin. In the large $U$ limit, electrons try to avoid double occupancy and thus each site becomes spin-polarized so as to form different long-range orders such as ferromagnetism, antiferromagnetism, ferrimagnetism, etc. \cite{GOUVEIA2015312,pssb2221560137} In the Schwinger boson representation \cite{PhysRevB.61.3508,PhysRevLett.96.016601}, the electron operator can be represented as
\begin{align}
c_{i\sigma}^\dagger = b_{i\sigma}^\dagger h_i+\sigma d_i^\dagger b_{i\bar{\sigma}},
\label{eq:Schwinger-boson-rep}
\end{align}
where $\sigma=\;\uparrow(+1),\downarrow(-1)$ is the spin index, $h_i$ and $d_i$ are the fermionic holon and doublon operators, $b_{i\sigma},b_{i\bar{\sigma}}$ are the Schwinger boson operators, and $\bar{\sigma}=-\sigma$ is the opposite spin of $\sigma$. By using the downfolding formula
\begin{align}
H_\mathrm{eff}=PHP-\frac{1}{U}PH\bar{P}HP+\mathcal{O}\!\left(\frac{1}{U^2}\right)\!,
\label{eq:downfolding-formula}
\end{align}
where $P$ is the projection operator into the Hilbert space of no doublons, we obtain the low-energy effective Hamiltonian of the chargeons
\begin{align}
H_\mathrm{eff}=\sum_{ij}\tilde{t}_{ij} h_ih_j^\dagger=\sum_{ij}\tilde{t}_{ij}f_i^\dagger f_j.
\label{eq:Ham-eff}
\end{align}
A particle-hole transformation has been done from the holons $h_i\mapsto f_i^\dagger$ to the chargeons, with the effective hopping amplitudes $\tilde{t}_{ij}$ given by
\begin{align}
\tilde{t}_{ij}=\sum_{\alpha\beta}
b_{i\alpha}^\dagger\!\left(t_{ij}^{\alpha\beta}
-\frac{1}{U}\sum_{k\gamma\delta}\gamma\delta\,t_{ik}^{\alpha\delta} t_{kj}^{\gamma\beta}\,b_{k\bar{\delta}}^\dagger b_{k\bar{\gamma}}\right)\!b_{j\beta}.
\label{eq:tij-eff}
\end{align}
The derivation of Eqs.~\eqref{eq:Ham-eff}--\eqref{eq:tij-eff} will be given in Appendix \ref{appendix:a}. For magnetically ordered systems, the bosonic operators can be viewed as $c$-numbers in the Bose-Einstein condensation (BEC) approximation \cite{PhysRevB.76.214428,davis2016formation}. Previous works on topological phase transitions mostly focus on those transitions induced by changes of the electronic hopping amplitudes $t_{ij}^{\alpha\beta}$, which may give rise to gap closing, band inversion \cite{PhysRevB.90.161202, Dongchao2017}, etc. Here with Eq.~\eqref{eq:tij-eff}, we can study two more types of topological phase transitions in terms of $\tilde{t}_{ij}$, i.e., a) those induced by changing the magnetic structure and b) those induced by $1/U$ that controls the strengths of the three-site virtual processes. This paper focuses on the latter situation. We consider gap closing of the chargeon bands induced by the change of Hubbard $U$ without changing the magnetic structure.

For simplicity, we consider a special case for 2D systems that the bare hopping $t_{ij}^{\alpha\beta}=t_{ij}^{\alpha}\delta_{\alpha\beta}$ conserves spin and that the magnetic structure is collinear ferrimagnetic in the $z$ direction. Since the magnetic moments have zero $x,y$ components, and no double occupancy is allowed in the large $U$ limit, every site can be occupied by either the spin-up electrons or the spin-down electrons only. In such a situation, Eq.~\eqref{eq:tij-eff} simplifies to
\begin{align}
\tilde{t}_{ij}=\sum_{\sigma}z_{i\sigma}^* z_{j\sigma}\left(t_{ij}^{\sigma}-\frac{1}{U}\sum_k t_{ik}^{\sigma}t_{kj}^{\sigma}|z_{k\bar{\sigma}}|^2\right)\!,
\label{eq:tij-eff-AFM}
\end{align}
where the bosonic operators $b_{i\sigma}\mapsto z_{i\sigma}$ have been mapped to $c$-numbers. Now we have a Hamiltonian with two sets of bands formed by electrons on spin-up sites and spin-down sites, which interact via the second-order virtual processes described by the three-site $\mathcal{O}(1/U)$ terms. In this paper, we use Eq.~\eqref{eq:tij-eff-AFM} as our simplified formula. Other magnetically ordered systems with more complex spin configurations such as noncollinear and spiral spin structures can be studied using Eq.~\eqref{eq:tij-eff}.

\section{Results \label{sec:results}}
\begin{figure}
\includegraphics[width=0.85\columnwidth]{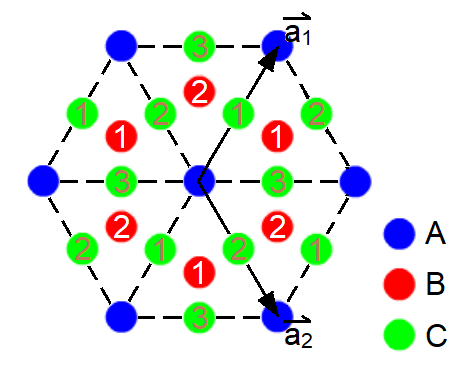}
\caption{The 2D hexagonal lattice structure $AB_2C_3$. The $A$ sites form a triangular lattice, the $B$ sites form a honeycomb lattice, and the $C$ sites form a Kagome lattice, all sharing the same lattice vectors $\vec{a}_1$ and $\vec{a}_2$.
\label{fig:AB2C3-lattice}}
\end{figure}

To study the topological phase transitions within the framework of Eq.~\eqref{eq:tij-eff-AFM}, we construct a 2D lattice structure $AB_2C_3$ with hexagonal symmetry (see Fig.~\ref{fig:AB2C3-lattice}). The $A$ sites form a triangular lattice with one band, which is topologically trivial. The $B$ sites form a honeycomb lattice with two bands, which realize the Haldane model. The $C$ sites form a Kagome lattice with three bands. We will put opposite magnetic moments on two of the three types of lattice sites and consider electronic phases in the $AB_2$, $AC_3$ and $B_2C_3$ structures, respectively.

\subsection{The AB$_2$ structure \label{sec:AB2-model}}
\begin{figure}
\includegraphics[width=\columnwidth]{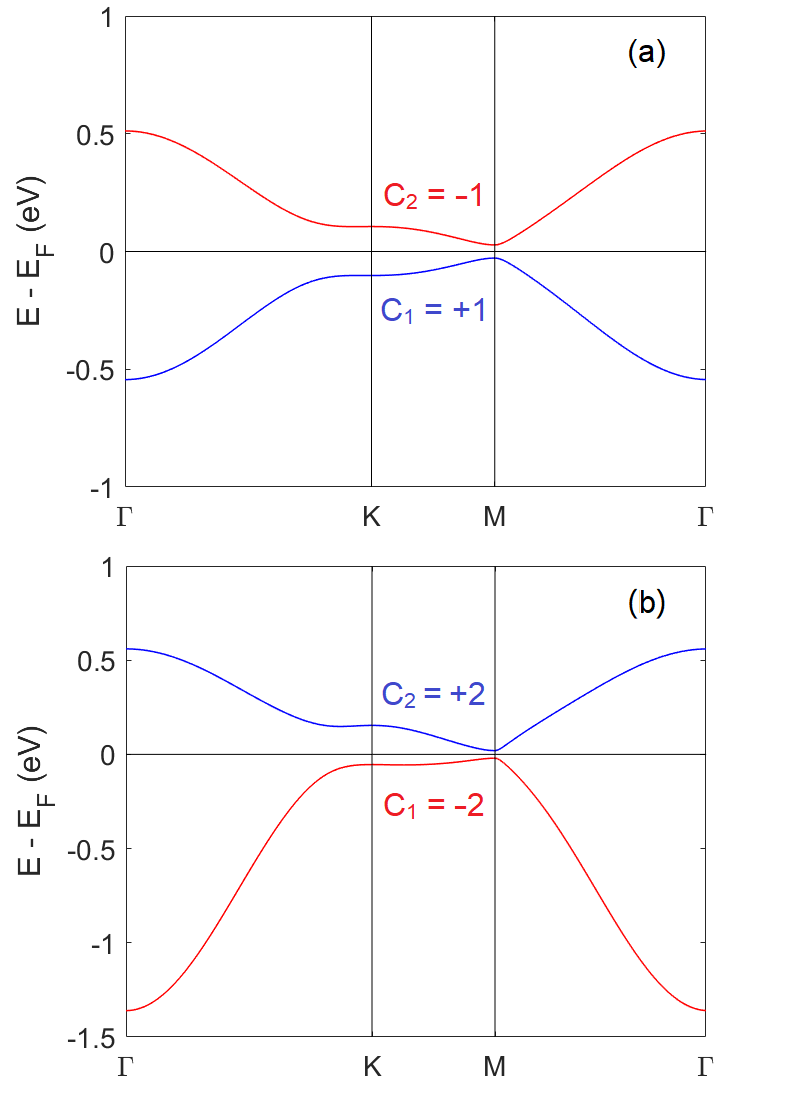}
\caption{The Chern bands of the $B$ sites (honeycomb) in the $AB_2$ model. Hopping amplitudes $t_1 = -0.15$~eV, $t_2 = (0.06 + 0.04i)$~eV, $t_3 = -0.01$~eV, $t = 0.8$~eV. Hubbard $U = 10$~eV in (a) and $U=4$~eV in (b). The Chern numbers $C_{1,2}$ indicate a topological phase transition (critical $U=5.3$~eV).
\label{fig:AB2-bands}}
\end{figure}
We consider an electronic phase with $N_{\uparrow}=N_{\downarrow}=1$ per unit cell. In case that the on-site orbital energy of an empty $A$ site is lower than that of an empty $B$ site, one of the spin species (e.g.~the $\downarrow$ electrons) would first singly occupy the $A$ sites. Then the other spin species (the $\uparrow$ electrons) would not occupy the $A$ sites because of the Hubbard $U$, but instead occupy the $B$ sites at occupancy $0.5$. When the SOC is considered, the $B$ sites become gapped and the $\uparrow$ electrons realize the Haldane model with real nearest-neighbor hopping $t_1$ and complex next-nearest-neighbor hopping $t_2$. We also consider a real para-position hopping $t_3$ among the $B$ sites and denote the real nearest $A$-$B$ site hopping as $t$. Following Eq.~\eqref{eq:tij-eff-AFM}, the effective hoppings $\tilde{t}_{1-3}$ are given by
\begin{align}
\tilde{t}_1=\frac{1}{2}\left(t_1-\frac{2t^2}{U}\right)\!,\quad
\tilde{t}_{2,3}=\frac{1}{2}\left(t_{2,3}-\frac{t^2}{U}\right)\!.
\end{align}
Here we assume the boson fields $z_{A\downarrow}=1$, $z_{A\uparrow}=0$ on the $A$ sites and $z_{B\uparrow}=1/\sqrt{2}$, $z_{B\downarrow}=0$ on the $B$ sites. In Eq.~\eqref{eq:tij-eff-AFM}, when the $i,j$ labels are on the $B$ sites, we have $\sigma=\;\uparrow$ and thus the $k$ label must be on the $A$ sites, which are occupied by $\bar{\sigma}=\;\downarrow$, so as to mediate a three-site virtual process $j\rightarrow k\rightarrow i$. All three hoppings $\tilde{t}_{1-3}$ are renormalized by such three-site virtual processes. Due to the three-site-enhanced hopping $\tilde{t}_3$, the $AB_2$ model can now realize beyond-Haldane phases with occupied-band Chern numbers $\pm 2$. 

In terms of the effective hoppings $\tilde{t}_{1-3}$, the spin-up Hamiltonian (i.e., chargeon Hamiltonian restricted to $B$ sites) in atomic gauge takes the form
\begin{align}
H_B(\vec{k})=\begin{bmatrix}
2\,_{\!}\mathrm{Re}\,_{\!}[\tilde{t}_2\zeta_2^*(\vec{k})] & \tilde{t}_1\zeta_1^*(\vec{k})+\tilde{t}_3\zeta_1(2\vec{k})
\vspace{0.5ex}\\
\tilde{t}_1\zeta_1(\vec{k})+\tilde{t}_3\zeta_1^*(2\vec{k}) & 2\,_{\!}\mathrm{Re}\,_{\!}[\tilde{t}_2\zeta_2(\vec{k})]
\end{bmatrix}\!,
\end{align}
where the functions $\zeta_{1,2}(\vec{k})$ are given by
\begin{subequations}
\begin{align}
\zeta_1(\vec{k})&=e^{i\vec{k}\cdot\frac{\vec{a}_1-\vec{a}_2}{3}}+e^{i\vec{k}\cdot\frac{\vec{a}_1+2\vec{a}_2}{3}}+e^{-i\vec{k}\cdot\frac{2\vec{a}_1+\vec{a}_2}{3}},\\
\zeta_2(\vec{k})&=e^{i\vec{k}\cdot\vec{a}_1}+e^{i\vec{k}\cdot\vec{a}_2}+e^{-i\vec{k}\cdot(\vec{a}_1+\vec{a}_2)}.
\end{align}
\end{subequations}
A topological phase transition can be realized as shown in Fig.~\ref{fig:AB2-bands}. In Fig.~\ref{fig:AB2-bands}a, the Hubbard $U=10$~eV is large. The two spin species are clearly separated by the Hubbard interaction with almost forbidden three-site virtual hoppings. The spin-up electrons form a Haldane phase on the $B$ sites with occupied-band Chern number $C_1=+1$ and unoccupied-band Chern number $C_2=-1$. The spin-down electrons fully occupy the triangular sites ($A$ sites) and form a topologically trivial band (not plotted) with Chern number $0$. As $U$ gets smaller, the three-site virtual processes $\sim\mathcal{O}(1/U)$ become stronger and the para-position hopping $\tilde{t}_3$ is significantly enhanced. The band gap in Fig.~\ref{fig:AB2-bands}a then closes at the M point at critical $U=5.3$~eV and reopens as $U$ is further reduced to form a beyond-Haldane phase with $C_1=-2$ and $C_2=+2$ (see Fig.~\ref{fig:AB2-bands}b for $U=4$~eV). Since the contribution $t^2/U$ of the second-order virtual processes is real, the imaginary part $\mathrm{Im}\,\tilde{t}_2=\mathrm{Im}\,t_2$ remains unaffected by $U$. Therefore, the system can undergo topological phase transitions between $C_1=+1\leftrightarrow -2$ (if $\mathrm{Im}\,t_2>0$) or $C_1=-1\leftrightarrow +2$ (if $\mathrm{Im}\,t_2<0$), but not in between the $C_1=\pm 1$ (or $\pm 2$) phases by tuning $U$.

\subsection{The AC$_3$ structure \label{sec:AC3-model}}
\begin{figure}
\includegraphics[width=\columnwidth]{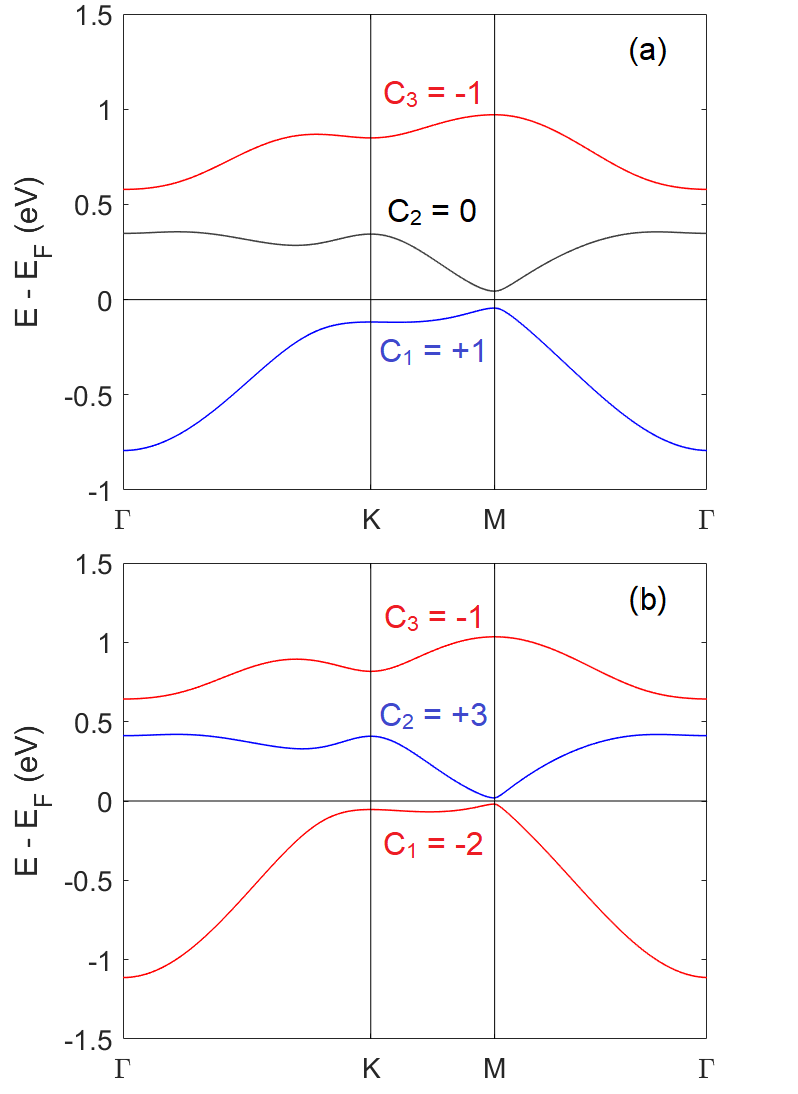}
\caption{The Chern bands of the $C$ sites (Kagome) in the $AC_3$ structure. Hopping amplitudes $t_1 = -(0.6+0.2i)$~eV, $t_2 = (0.1 + 0.1i)$~eV, $t_3 = -0.25$~eV, $t = 0.8$~eV. Hubbard $U = 10$~eV in (a) and $U=4$~eV in (b). The Chern numbers $C_{1-3}$ indicate a topological phase transition (critical $U=4.9$~eV).
\label{fig:AC3-bands}}
\end{figure}
Consider an electronic phase in which the $A$ sites are singly occupied by the $\downarrow$ electrons and the $C$ sites are occupied by the $\uparrow$ electrons at occupancy $1/3$. The situation is similar to $AB_2$, except that the $C$ sites form a Kagome lattice. We consider the nearest-neighbor and next-nearest-neighbor hoppings $t_1$, $t_2$, and real para-position hopping $t_3$ of the $C$-site hexagons. Both $t_1$ and $t_2$ can be complex. The real nearest-neighbor $A$-$C$ site hopping is denoted as $t$. From Eq.~\eqref{eq:tij-eff-AFM}, we have
\begin{align}
\tilde{t}_{1-3}=\frac{1}{3}\left(t_{1-3}-\frac{t^2}{U}\right)\!,
\end{align}
assuming $z_{A\downarrow}=1$, $z_{A\uparrow}=0$ for the $A$ sites and $z_{C\uparrow}=1/\sqrt{3}$, $z_{C\downarrow}=0$ for the $C$ sites. In terms of $\tilde{t}_{1-3}$, the $C$-site Kagome Hamiltonian takes the form
\begin{align}
H_C(\vec{k})=\sum_{\nu=1}^3 H_C^{(\nu)}(\vec{k}),
\label{eq:Kagome-Ham}
\end{align}
where the nearest-neighbor hopping $H_C^{(1)}(\vec{k})$, the next-nearest-neighbor hopping $H_C^{(2)}(\vec{k})$ and the para-position hopping $H_C^{(3)}(\vec{k})$ Hamiltonians are given specifically in Appendix \ref{appendix:b}. A topological phase transition analogous to the $AB_2$ situation is realized in Fig.~\ref{fig:AC3-bands}. In Fig.~\ref{fig:AC3-bands}a, the Hubbard $U=10$~eV is large and the three-site virtual hoppings are almost forbidden. As $U$ gets smaller, the para-position hopping $\tilde{t}_3$ is significantly enhanced. The occupied-band Chern number changes from $C_1=+1$ (see Fig.~\ref{fig:AC3-bands}a) to $C_1=-2$ (see Fig.~\ref{fig:AC3-bands}b) when the gap closes at the M point at critical $U=4.9$~eV. In the mean time, the Chern number $C_2$ of the middle band changes from $0$ to $+3$ and the Chern number of the flat band on the top $C_3=-1$ remains unchanged.

In Secs.~\ref{sec:AB2-model}--\ref{sec:AC3-model}, we have studied the enhancement effect of para-position hopping $\tilde{t}_3$ due the three-site virtual processes proportional to $1/U$. We find that in both the honeycomb and the Kagome lattices, the three-site processes can lead to topological phase transitions of $C_1=+1\leftrightarrow-2$ (or symmetrically $C_1=-1\leftrightarrow+2$) by closing the band gap at the M point. Because $t^2/U$ is real, we cannot realize topological phase transitions between the $C_1=\pm 1$ phases. We will demonstrate in Sec.~\ref{sec:B2C3-model} that the $+1\leftrightarrow-1$ transitions can be realized in the B$_2$C$_3$ model by making the contributions of the three-site processes $\mathcal{O}(tt'/U)$ complex.

\subsection{The B$_2$C$_3$ structure \label{sec:B2C3-model}}
\begin{figure}
\includegraphics[width=0.9\columnwidth]{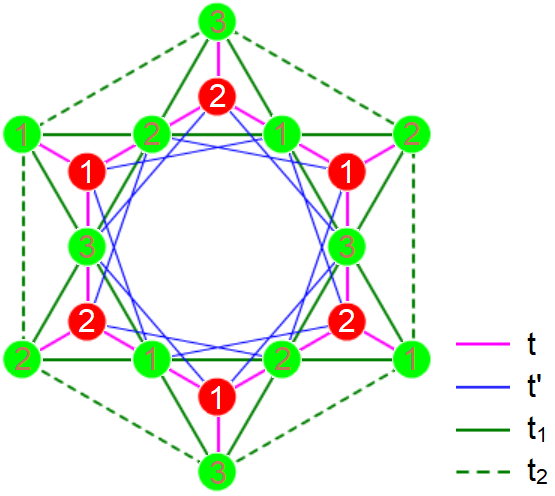}
\caption{The hoppings considered in B$_2$C$_3$. Here $t,t'$ are between the $B$-$C$ sites and $t_1,t_2$ are the nearest-neighbor and next-nearest-neighbor hoppings of the $C$-site Kagome lattice. All hoppings except $t$ can be complex due to the SOC. The para-position hoppings are ignored.
\label{fig:B2C3-hoppings}}
\end{figure}
\begin{figure}
\includegraphics[width=\columnwidth]{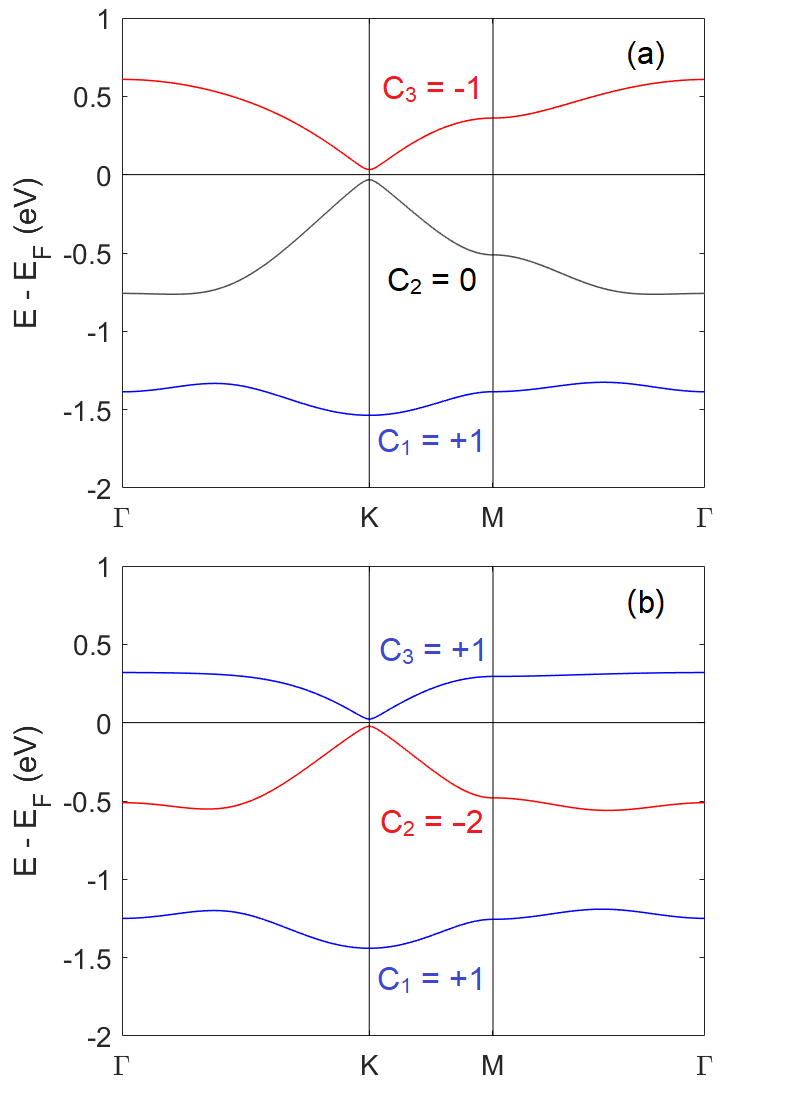}
\caption{The Chern bands of the $C$ sites in the $B_2C_3$ model. Hopping amplitudes $t_1 = (0.6-0.1i)$~eV, $t_2 = -(0.1 + 0.02i)$~eV, $t = 0.8$~eV, $t'=(0.1+0.1i)$~eV. Hubbard $U = 10$~eV in (a) and $U=4$~eV in (b). The Chern numbers $C_{1-3}$ indicate a topological phase transition (critical $U=5.3$~eV).
\label{fig:B2C3-bands}}
\end{figure}

In this section, we consider an electronic phase with $N_{\uparrow}=N_{\downarrow}=2$ per unit cell. Let the two $B$ sites in a unit cell be singly occupied by the $\downarrow$ electrons and the three $C$ sites be occupied by $\uparrow$ electrons at occupancy $2/3$. We consider the hoppings $t,t'$ between the $B$-$C$ sites and hoppings $t_1,t_2$ among the $C$ sites as shown in Fig.~\ref{fig:B2C3-hoppings}. Since the total Chern number of the two spin-down bands on the $B$ sites is zero (c.f.~Fig.~\ref{fig:AB2-bands}a), we focus on the topological properties of the Kagome bands, which are controlled by $1/U$. From Eq.~\eqref{eq:tij-eff-AFM}, we have
\begin{align}
\tilde{t}_1=\frac{2}{3}\left(t_1-\frac{t^2}{U}\right)\!,\quad
\tilde{t}_2=\frac{2}{3}\left(t_2-\frac{2tt'}{U}\right)\!,
\label{eq:t-tilde-B2C3}
\end{align}
assuming $z_{B\downarrow}=1$, $z_{B\uparrow}=0$ and $z_{C\uparrow}=\sqrt{2/3}$, $z_{C\downarrow}=0$. The para-position hopping $\tilde{t}_3=0$ of the Kagome lattice is ignored. Even though $\tilde{t}_3$ can be mediated by $t'^2/U$, these contributions are small assuming $|t|\gg |t'|$. Notice that the $B$-$C$ site hopping $t$ is real, while $t'$ can be complex due to SOC. We define for $\uparrow$ electrons that the blue-line hoppings in Fig.~\ref{fig:B2C3-hoppings} is $t'$ in clockwise directions and $(t')^*$ in counter-clockwise directions. The Hamiltonian $H_C(\vec{k})$ is still given in Appendix \ref{appendix:b} with the effective hoppings $\tilde{t}_{1,2}$ now given by Eq.~\eqref{eq:t-tilde-B2C3}. A topological phase transition is realized as shown in Fig.~\ref{fig:B2C3-bands}.

In Fig.~\ref{fig:B2C3-bands}a, the Hubbard $U=10$~eV and the occupied-band Chern number $C_1+C_2=+1$, which is determined by the imaginary parts $\mathrm{Im}\,\tilde{t}_{1,2}$ of the effective hoppings in the Kagome lattice. As $U$ gets smaller, since $t^2/U$ is real, $\mathrm{Im}\,\tilde{t}_1$ remains unchanged, so only the $tt'/U$ term in Eq.~\eqref{eq:t-tilde-B2C3} can affect $\mathrm{Im}\,\tilde{t}_2$. The band gap closes at the K point at critical $U=5.3$~eV and then reopens to give rise to a $C_1+C_2=-1$ phase as $U$ further decreases to $4$~eV (see Fig.~\ref{fig:B2C3-bands}b). The Chern number of the flat band at the bottom $C_1=+1$ remains unchanged throughout the process. Because the imaginary part of the hopping amplitudes can be tuned by $1/U$, the phase separation between Chern numbers $\pm 1$ is broken. A topological phase transition between the $\pm 1$ phases can now be realized by tuning the Hubbard $U$ due to the complex virtual hopping $\mathcal{O}(tt'/U)$.

\section{Conclusion \label{sec:conclusion}}
We have demonstrated in this paper that the three-site virtual processes in the large $U$ limit of the Hubbard model can exhibit interesting renormalization effects of the hopping amplitudes and give rise to topological phase transitions in the low-energy effective theory. We constructed 2D lattice models to realize the $1/U$ control of the honeycomb and Kagome lattices. In the $AB_2$ model, a topological phase transition between the Haldane phase and beyond-Haldane phase is realized by considering the enhancement effect of the para-position hopping $\tilde{t}_3$ due to the $A$-site mediated virtual hoppings proportional to $1/U$. The $AC_3$ model realizes a similar phase transition on the Kagome lattice. Both transitions close the band gap at the M point. In the $B_2C_3$ model, we also realize topological phase transitions on the Kagome lattice, but the band gap closes at the K point. The contribution $\mathcal{O}(tt'/U)$ of the three-site processes can be complex and drives the system across the phases boundary of occupied-band Chern number $=\pm 1$.

The phase transitions found in our model studies are realized using collinear antiferromagnetic (or ferrimagnetic) spin configurations. The spin-up and spin-down electrons occupy inequivalent lattice sites. In the examples shown in this paper, for simplicity, we let one spin species fully occupy one type of lattice site so as to be topologically trivial, and use them to control the topological phase of the other spin species via the three-site terms. Interesting directions for further studies could be that both spin species exhibit topological properties and mutually influence each other via the three-site terms, or the realization of similar Coulomb engineering effects in non-collinear spin systems.

\begin{figure}[t!]
\includegraphics[width=\columnwidth]{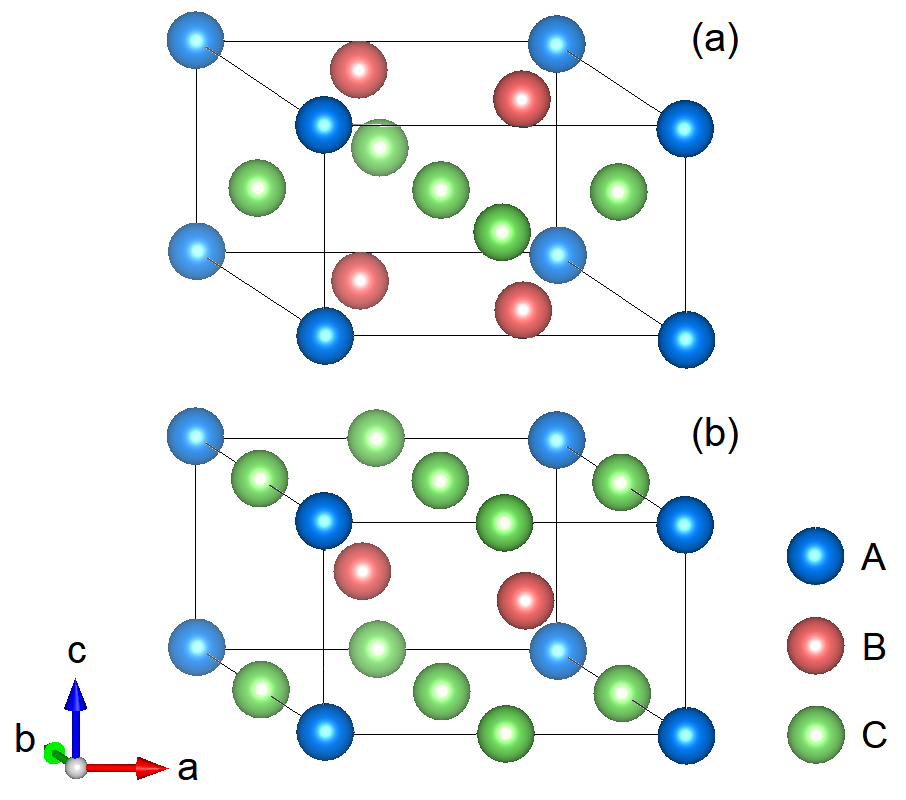}
\caption{Possible realizations of the $AB_2C_3$ lattice in a 3D hexagonal crystal structure with (a) alternating $AB_2$ and $C_3$ layers and (b) alternating $AC_3$ and $B_2$ layers. Both structures have the P6/mmm space group symmetry.
%The structure in Panel (a) is more commonly seen in ICSD crystals. Variations and unit-cell doubling of Panel (a) are also realizable.
\label{fig:3D-motifs}}
\end{figure}

Finally, we would like to discuss the possible realizations of our model in real materials. The $1/U$-controlled topological phase transitions can be realized without restricting the atoms to the same 2D plane. Two possible 3D structures are shown in Fig.~\ref{fig:3D-motifs}, both with P6/mmm symmetry. Examples of materials with the structure of Fig.~\ref{fig:3D-motifs}a are $R$Co$_3$B$_2$ \cite{doi:10.1063/1.370047} with $R=\,$rare-earth elements, and also GdNi$_3$Ga$_2$ \cite{PECHEV200062}, etc, which are potential candidates for the $AC_3$ model of two types of magnetic atoms. Coplanar $AC_3$ candidates include the TiNi$_3$-type compounds \cite{PhysRevB.48.4276, WODNIECKA1995128, PhysRevB.82.155136} with shifted layers of close-packed $AC_3$ structures. Candidates for the $AB_2$ model include e.g.~UNi$_2$Al$_3$ \cite{GEIBEL1993188} and EuCo$_2$Al$_9$ \cite{ThiedeJeitschko}, etc. %The $B_2C_3$ model with two types of magnetic atoms are a bit rare, but interestingly, the same structure with coplanar $B$ and $C$ atoms is seen in EuCo$_2$Al$_9$ \cite{ThiedeJeitschko} with $B=\mathrm{Co}$ and $C=\mathrm{Al}$.
We expect our work to be interesting to the fields of magnetism in alloys, ferrimagnets and other materials with multiple types of magnetic atoms.

\section*{Acknowledgments}
This work is supported by the National Key Research and Development Program of China (2018YFA0307000), and the National Natural Science Foundation of China (11874022). Z. H would thank the support of the 66th Chinese Postdoc Fellowship. We would also like to thank the helpful discussions with Prof.~Biao Lian at Princeton Center for Theoretical Science
in Princeton University at the early stage of this work.
%regarding the Schwinger boson technique and topological phase transitions.

\begin{widetext}
\appendix
\section{Derivation of the low-energy effective Hamiltonian \label{appendix:a}}
By plugging Eq.~\eqref{eq:Hubbard-model} into Eq.~\eqref{eq:downfolding-formula}, one obtains
\begin{align}
H_\mathrm{eff}&=\sum_{ij\alpha\beta} t_{ij}^{\alpha\beta}Pc_{i\alpha}^\dagger c_{j\beta}P-\frac{1}{U}\sum_{ijkl}\sum_{\alpha\beta\gamma\delta}t_{ij}^{\alpha\beta}t_{kl}^{\gamma\delta} Pc_{i\alpha}^\dagger c_{j\beta}\bar{P}c_{k\gamma}^\dagger c_{l\delta}P.
\end{align}
Then plugging in Eq.~\eqref{eq:Schwinger-boson-rep}, one finds that the projections $P$ pick out the following terms
\begin{align}
H_\mathrm{eff}=\sum_{ij\alpha\beta}t_{ij}^{\alpha\beta} h_ih_j^\dagger b_{i\alpha}^\dagger b_{j\beta}-\frac{1}{U}\sum_{ijkl}\sum_{\alpha\beta\gamma\delta}\beta\gamma\,t_{ij}^{\alpha\beta}t_{kl}^{\gamma\delta}\,b_{i\alpha}^\dagger b_{j\bar{\beta}}^\dagger\,b_{k\bar{\gamma}}\,b_{l\delta}\,h_i\,(Pd_jd_k^\dagger P)\,h_l^\dagger.
\end{align}
The bosonic operators are automatically normal ordered. Notice that $Pd_jd_k^\dagger P=\delta_{jk}$, because the doublon created must also be the doublon destructed so as to go back to the no-doublon subspace. One may then set $j=k$ and rename the dummy indices $l\mapsto j$ and $\beta\leftrightarrow\delta$ to obtain
\begin{align}
H_\mathrm{eff}=\sum_{ij}h_ih_j^\dagger\left[\sum_{\alpha\beta}b_{i\alpha}^\dagger\!\left(t_{ij}^{\alpha\beta}-\frac{1}{U}\sum_{k}\sum_{\gamma\delta}\gamma\delta\,t_{ik}^{\alpha\delta}t_{kj}^{\gamma\beta}\,b_{k\bar{\delta}}^\dagger\,b_{k\bar{\gamma}}\right)\!b_{j\beta}\right]\!.
\end{align}
This result agrees with Eqs.~\eqref{eq:Ham-eff}--\eqref{eq:tij-eff} in the main text by defining the quantity in the square bracket as $\tilde{t}_{ij}$. All $\mathcal{O}(1/U)$ renormalizations of $\tilde{t}_{ij}$ are considered in this formalism. Then we do a particle-hole transformation $h_i\mapsto f_i^\dagger$ to the holon operators and map the bosonic operators $b_{i\sigma}\mapsto z_{i\sigma}$ to $c$-numbers and obtain
\begin{align}
H_\mathrm{eff}=\sum_{ij}f_i^\dagger f_j\left[\sum_{\alpha\beta}z_{i\alpha}^* z_{j\beta}\left(t_{ij}^{\alpha\beta}-\frac{1}{U}\sum_{k\gamma\delta}\gamma\delta\,t_{ik}^{\alpha\delta}t_{kj}^{\gamma\beta}\,_{\!}z_{k\bar{\delta}}^* z_{k\bar{\gamma}}\right)\right]=\sum_{ij}\tilde{t}_{ij}f_i^\dagger f_j.
\end{align}
In the special case that the bare hopping $t_{ij}^{\alpha\beta}=t_{ij}^{\alpha}\delta_{\alpha\beta}$ conserves spin, we have
\begin{align}
\tilde{t}_{ij}=\sum_{\alpha\beta}z_{i\alpha}^* z_{j\beta}\left(t_{ij}^{\alpha}\delta_{\alpha\beta}-\frac{1}{U}\sum_{k}\alpha\beta\,t_{ik}^{\alpha} t_{kj}^{\beta} z_{k\bar{\alpha}}^* z_{k\bar{\beta}}\right)\!.
\end{align}
Then the collinear ferrimagnetic structure in the $z$ direction (perpendicular to the 2D lattice plane) with no double occupancy eliminates the $\alpha\neq\beta$ terms because site $k$ can only be occupied by one type of spin species. Therefore, one obtains Eq.~\eqref{eq:tij-eff-AFM} in the main text.

\section{Kagome Hamiltonian in terms of $\tilde{t}_{1-3}$ \label{appendix:b}}
In terms of the effective hoppings $\tilde{t}_{1-3}$, the full Kagome Hamiltonian $H_C(\vec{k})$ contains 3 parts as defined by Eq.~\eqref{eq:Kagome-Ham}: the nearest-neighbor hopping Hamiltonian as given by
\begin{align}
H_C^{(1)}(\vec{k})=\begin{bmatrix}
0 & 2\tilde{t}_1\cos\left(\vec{k}\cdot\frac{\vec{a}_1+\vec{a}_2}{2}\right) & 2\tilde{t}_1^*\cos\left(\vec{k}\cdot\frac{\vec{a}_2}{2}\right)\\
2\tilde{t}_1^*\cos\left(\vec{k}\cdot\frac{\vec{a}_1+\vec{a}_2}{2}\right) & 0 & 2\tilde{t}_1\cos\left(\vec{k}\cdot\frac{\vec{a}_1}{2}\right)\\
2\tilde{t}_1\cos\left(\vec{k}\cdot\frac{\vec{a}_2}{2}\right) & 2\tilde{t}_1^*\cos\left(\vec{k}\cdot\frac{\vec{a}_1}{2}\right) & 0
\end{bmatrix}\!,
\end{align}
the next-nearest-neighbor hopping Hamiltonian as given by
\begin{align}
H_C^{(2)}(\vec{k})=\begin{bmatrix}
0 & 2\tilde{t}_2\cos\left(\vec{k}\cdot\frac{\vec{a}_1-\vec{a}_2}{2}\right) & 2\tilde{t}_2^*\cos\left[\vec{k}\cdot\left(\vec{a}_1+\frac{\vec{a}_2}{2}\right)\right]\\
2\tilde{t}_2^*\cos\left(\vec{k}\cdot\frac{\vec{a}_1-\vec{a}_2}{2}\right) & 0 & 2\tilde{t}_2\cos\left[\vec{k}\cdot\left(\frac{\vec{a}_1}{2}+\vec{a}_2\right)\right]\\
2\tilde{t}_2\cos\left[\vec{k}\cdot\left(\vec{a}_1+\frac{\vec{a}_2}{2}\right)\right] & 2\tilde{t}_2^*\cos\left[\vec{k}\cdot\left(\frac{\vec{a}_1}{2}+\vec{a}_2\right)\right] & 0
\end{bmatrix}\!,
\end{align}
and the para-position hopping Hamiltonian as given by
\begin{align}
H_C^{(3)}(\vec{k})=\begin{bmatrix}
2\tilde{t}_3\cos\left(\vec{k}\cdot\vec{a}_1\right) & 0 & 0\\
0 & 2\tilde{t}_3\cos\left(\vec{k}\cdot\vec{a}_2\right) & 0\\
0 & 0 & 2\tilde{t}_3\cos\left[\vec{k}\cdot(\vec{a}_1+\vec{a}_2)\right]
\end{bmatrix}\!,
\end{align}
all written in the atomic gauge. In the $AC_3$ model, we consider the $1/U$-control of all three effective hoppings $\tilde{t}_{1-3}$. In the B$_2$C$_3$ model, we restrict ourselves to $\tilde{t}_{1-2}$, i.e., setting $\tilde{t}_3=0$.
\end{widetext}

\bibliography{Hubbard}
\end{document}